\documentstyle[aps,preprint,eqsecnum]{revtex} \baselineskip 12pt
\begin{document}

\title{WT IDENTITIES FOR PROPER VERTICES AND RENORMALIZATION IN A
 SUPERSPACE FORMULATION OF GAUGE THEORIES}

\vspace{.7in}

\author{ Satish D. Joglekar\footnote{e-mail address:- sdj@iitk.ernet.in }}

\address{Department of physics \\ Indian Institute of Technology, Kanpur\\
Kanpur, 208016, (INDIA)\\ and}

\author{Bhabani Prasad Mandal\footnote{ Address for correspondence,    e-mail
:- bpm@iopb.ernet.in}}

\address{
Institute of Physics, Sachivalaya Marg,\\ Bhubaneswar-751005, India,\\
}

\vspace{.7in}

\maketitle

\vspace{.8in}

\begin{abstract}
We formulate the WT identity for proper vertices in a simple and
compact form $ \frac{ \partial \Gamma }{\partial \theta } =0 $ in a
superspace formulation of gauge theories proposed earlier. We show
 this WT identity ( together with a subsidiary constraint) lead, in
transparent way, the superfield superspace  multiplet renormalizations
formulated  earlier  (~and shown to explain symmetries of Yang-Mills theory 
renormalization ). 
\end{abstract}
\newpage

\section{INTRODUCTION}
Yang-Mills theories have acquired a central place in the theoretical
formulation of strong, weak and electro-magnetic interactions ( the
Standard Model) \cite{cheng}. Central to Yang-Mills theories is the
local gauge invariance of the basic action. This basic gauge invariance 
manifests, in particular, as the global BRS invariance of the effective
action \cite{bsr}.
The  consequences of BRS invariance, formulated as WT identities,
are central to the discussion of renormalizability, unitarity, gauge
independence of the theory \cite{lee,pr1}. Any attempt that sheds light on,
offer a reformulation of, and leads to simplified understanding of BRS 
symmetry and  Yang-Mills theory is therefore of significance to
particle physics.

With the above aim in mind, viz. to simplify the expression of BRS
invariance, of WT identities and to simplify the discussion
of renormalization of gauge theories,
a superspace formulation of gauge theories was proposed \cite{sdj}.
It was an improvement upon a number of earlier superfield /superspace 
formulations \cite{six,sev,pr2} in a number of ways. Firstly, unlike the earlier
formulations where structure of superfields was, in effect, determined
by hand, the superfield here \cite{sdj} had all their components free. 
The $6$- dimensional superspace was also one in which superspace 
rotations could be freely carried out [unlike Ref. 7]. BRS transformations 
were effectively generated from within. The sources for fields and BRS
composite operators aroze out of a multiplet sources structure. This
has been summarized in Ref. 5.

The superspace formulation of Ref. 5 was further developed in a number of
ways. It was shown that the WT identities in Yang-Mills theories could
be cast in a neat and compact form,
$ \frac{ \partial \bar{W} }{ \partial \theta } =o $
\cite{sdj1}. It was further shown that this compact form can be derived
by simply considering an $OSp(3,1|2)$ rotation in superspace which is
an approximate symmetry of the superspace generating functional
$ \bar{W}$ \cite{osp}. Thus in Ref. 10 it was shown that, a \underline{
coordinate rotation} in superspace contains all the consequences of the
\underline{field transformation} ( BRS ). It was later shown that this
coordinate rotation is equivalent to a set of local BRS transformations \cite{local}.
The superspace  
of Ref. 5 was generalized to include  the anti- BRS symmetry \cite{zpc}
scalars and  gauge invariant operators \cite{D52}.

Of particular interest to the present work are results of Ref. 14, where it was shown
that renormalization transformations in Yang-Mills theories take a simple
form when expressed in the superspace formalism. Interrelations between
renormalizations of all sources for fields and composite operators is shown to
arize from the fact that they belong together in source supermultiplets which
transform as whole under renormalization . The OSp-symmetry breaking piece
of the Lagrangian, ${\cal L}_1$ 
( that generates a large numbers of terms: sources for the fields
\underline{and} composite operators, gauge-fixing and ghost term)
was shown to be form invariant under renormalization transformations.

One drawback of the result of Ref. 14, despite their attractiveness, has
been that they were not directly `` derived" from the superspace WT
identity but shown to be true by invoking the correspondence between
superspace generating functional and the ordinary Yang-Mills theory
established
in Ref. 5. The main aim of this work is to fill up this technical gap and
complete the derivation of the superspace renormalization of gauge
theories.

A natural tool for discussing renormalization is the WT identities
for proper vertices. This was not done in Ref. 9 and 10, where the aim was
to formulate the WT identity for $ \bar{W} $
in a simple form. As done in section II of this work sources now have to be
introduced for \underline{all} elementary fields (including of those called
$A_{i, \lambda }$ ) and this was not necessary in Ref. 9, 10. These
modifications allow one to Legendre transform and cast the WT identity for
$ \Gamma $ in a equally simple form $
\frac{ \partial \Gamma }{ \partial \theta }=0$ 
(section III). This simplifies the discussion of renormalization of gauge
theories as seen in section IV. The simplification is seen to arize from
the fact that the nilpotent operator $\frac{ \partial }{\partial \theta }$
independent of the variables (fields, coupling constant etc) that get
renormalized.

\section{Preliminary}

In this section, we shall introduce our notations, review past results
and introduce slightly modified generating functional of superspace
formulation .

\subsection{The Superspace Formulation }

We shall work in the context of a pure Yang-Mills theory with simple
gauge group given by the Lie algebra of generators.
\begin{equation}
\left [ T^\alpha ,T^\beta \right ] = if^{\alpha \beta \gamma } T^\gamma 
\label{21}
\end{equation}
and the covariant derivative
\begin{equation}
D_\mu^{\alpha \beta} c^\beta = \left ( - \partial _\mu \delta ^{\alpha \beta}
+gf^{\alpha \beta \gamma } A^\gamma _\mu \right )c^\beta
\label{22}
\end{equation}
$f^{\alpha \beta \gamma } $ are totally anti-symmetric.

The generating functional for the Green's functions in linear gauges,
( with additional sources for BRS variations introduced ) is given by
\begin{eqnarray}
W &=& \int [ dA\, dc \, d\zeta ] \exp\left(iS +i\int d^4x\left\{j^\alpha _\mu
A^{\alpha \mu}+\bar{\xi}^\alpha c^\alpha +\zeta^\alpha \xi^\alpha
+\kappa^{\alpha \mu } D^{\alpha \beta}_\mu c^\beta + \frac{g}{2}l^\alpha 
f^{\alpha \beta \gamma }c^\beta c^\gamma \right \}\right)\label{225}\nonumber\\
 &&\equiv W[j,\bar{\xi},\xi ,\kappa, -l, t]
\label{23}
\end{eqnarray}
Where the action $S$ is given by,
\begin{equation}
S= S_0+ S_g + S_{gf}
\label{24}
\end{equation}
With
\begin{eqnarray}
S_0 &=& \int d^4x \left\{ - \frac{ 1}{4} F^\alpha
_{\mu\nu}F^{\alpha\mu\nu}\right\}\nonumber\\
S_g &=& \int d^4x \left\{-\partial ^\mu\zeta^\alpha D^{\alpha \beta}_\mu
c^\beta\right\}\nonumber\\
S_{gf} &=& \int d^4x \left \{ - \frac{ 1}{2\eta_0} (\partial \cdot A^\alpha
+t^\alpha )^2\right\}
\label{25}
\end{eqnarray}

We shall briefly introduce the superspace formulation first introduced
in Ref. 5. It was a six dimensional superspace which has two anti-commuting
dimensions $ \lambda $ and $\theta $ such that the 6-dimensional coordinate
vector is $\bar{x} ^i\equiv (x^\mu, \lambda ,\theta )$. $\lambda $ and
$\theta $ are scalars under the Lorentz transformations . In this space, a
metric is introduced whose non vanishing components are $g_{00} = -g_{11}
=-g_{22}= -g_{33}= -g_{45} =g_{54}=1$. The group of linear homogeneous
transformations that preserves $ \bar{x} ^ig_{ij}\bar{x} ^j$ is
$OSp(3,1|2)$.

The superspace formulation of Ref. 5 introduces superfields $\bar{A}^\alpha _i
(\bar{x} ) $ and $ \zeta^\alpha (\bar{x} ) $ transforming as a covariant
vector and a scalar under $ OSp(3,1|2)$. The superfield $\bar{A}^\alpha _\mu
(\bar{x} ) $ contains a component  $ A^\alpha _\mu (x) $ which is identified with the
the usual Yang-Mills field; $ A^\alpha _5(x) \equiv c^\alpha _5 (x) $
is identified with the usual ghost field and $\zeta^\alpha (x)$
with anti-ghost field . The remaining fields in the expression
of $A^\alpha _i (\bar{x} ) $ and $ \zeta^\alpha (\bar{x} )$
in terms of $ \lambda $ and $ \theta $ are certain auxiliary 
fields whose role is primarily to generate the BRS constraints from within
the superfield action. We also introduce  two vector supersources 
$K^{\alpha i}( \bar{x} ), L^{\alpha i}( \bar{x} )$ and a scalar supersource
$t^\alpha (\bar{x} )$. As was shown in Ref. 5, the source $K^{\alpha i}(\bar{x}
)$ contains compactly in it sources for the gauge field , the ghost field 
and also the sources for the BRS variation composite operators ( $\kappa $
and $l$) all in a single supermultiplet of $OSp(3,1|2)$; and a similar
statement holds for $t( \bar{x} )$.

In this work, we use a superspace formulation which is somewhat modified
compared to Ref.5 . We introduce the superspace action
\begin{equation}
\bar{S} = \int d^4 x \left\{ - \frac{ 1}{4} g^{ik}g^{jl}F^\alpha
_{ij}F^\alpha_{kl}\right\} + S_1 \equiv S_0 +S_1
\label{26}
\end{equation}
with the symmetry breaking piece,
\begin{equation}
S_1= \int d^4x \left [\frac{  \partial 
}{\partial \theta} \left\{ \bar{K}^{\alpha i} \bar{A}^\alpha _i +\zeta^\alpha \left[
\partial ^\mu A^\alpha _\mu (\bar{x} ) + \frac{ 1}{2\eta_0} \zeta_{,\theta }^\alpha 
(\bar{x} ) + t^\alpha (\bar{x} )\right]\right\} + \frac{ \partial  }{ \partial \lambda }
[\bar{L} ^{\alpha i}\bar{A}^\alpha _i] + \partial \cdot L\zeta \right ]
\label{27}
\end{equation}
and $F_{ij}^\alpha $,  superspace field strength tensor, is defined as 
\begin{equation}
F^\alpha_{ij} \equiv \partial_i\bar{A}^\alpha_j(\bar{x}) - \bar{A}^\alpha_i(\bar{x})
\stackrel{\leftarrow}\partial_j + g F^{\alpha\beta\gamma} \bar{A}^\beta_i(\bar{x})\bar{A}^\gamma(\bar{x})
\nonumber
\end{equation}
where $i,j$ takes values $0,1 \cdots 5$.

We, further, introduce the generating functional
\begin{equation}
\bar{W}[\bar{K} ,\bar{L} ,t] =\int \{d\bar{A}\}\{d\zeta\} e^{i\bar{S  
}[ A,\zeta, \bar{K} , \bar{L} ,t]}
\label{28}
\end{equation}
Where the measure is defined as
\begin{eqnarray}
\left\{ d\bar{A}\right\} &=& \left\{ dA\right\}\left\{dc_4\right\}\left\{dc_5\right\}
\nonumber \\
\left\{dA\right\} &=& \prod _{\alpha\mu x} dA^\alpha_\mu(\bar{x})\,\, dA^\alpha_{\mu,\lambda}
(\bar{x}) \,\, dA^\alpha_{\mu,\theta}(\bar{x});\,\,\, \,\,\,\,\,\, 0\leq\mu\leq 3\nonumber 
\end{eqnarray}
and similar definitions for $ \left\{dc_4\right\}$ and $\left\{ dc_5\right\}$.

The generating functional in \ref{28} differs from the one in
Ref. 5 only by the addition of a new sources $ \bar{L}^{\alpha i}(\bar{x} )
  $ in \ref{27}. This source has been added to provide a source for the
field $A_{i,\lambda }$ [ this is necessary before a Legendre transformation
can be carried out to generate $\Gamma $ , as a source for every field is
needed]. The last term in $S_1$ viz. $\partial \cdot L \zeta $ is added not so
much out of necessity, but for the sake of certain anticipated simplifications.

We note that it is enough to consider $\bar{W} $ at $ L^i_{, \lambda } =0=
L^i_{, \theta } $ because $L^i_{, \lambda }$ is only a part of a source for
$A_i$ ( and hence can be dropped) and $L^i_{, \theta } $ is not a source
for any field. Hence, in future we shall assume that 
$L^i_{, \lambda } =0 = L^i_{, \theta }.$

\subsection{Review of  some of the earlier results}

Here, we review some of the relevant earlier results. The generating
functional of Eq.\ref{28} , at $L=0$, has been shown to be related to the
generating functional of Yang-Mills theory of Eq.\ref{23} via \cite{sdj}
\begin{equation}
\int [dK^4][dK^4_{, \theta }] \bar{W} [\bar{K} , \bar{L} ,t]\left |_{L=0} = 
W\left [ K^{\alpha \mu}_{, \theta };K^{\alpha 5}_{, \theta };-t^\alpha _{,
\theta };K^{\alpha \mu};K^{\alpha 5}; t^\alpha \right ]\right . 
\label{29}
\end{equation}
The WT identities of gauge theories have been cast in a simple form, at
$L=0$ \cite{sdj1}
\begin{equation}
\frac{ \partial \bar{W} }{\partial \theta } = 0
\label{30}
\end{equation}
The underlying Lagrangian possesses broken $OSp(3,1|2)$ invariance in the
superspace. It has been shown that a superrotation in superspace takes the
place of BRS transformations in this formulation and this has been used to
derived \ref{30} directly \cite{osp}.

It has been shown \cite{D49} that the renormalization transformations of sources
for fields \underline{and} composite BRS variation operators, when cast in the superspace formulation
[using the correspondence in Eq. \ref{29} and \ref{23}], take a very simple
form. They consist of (i) a supermultiplet renormalization as a whole (ii)
renormalizations of $\lambda $ and $\theta $ (iii) an optional $Sp(2)$ rotation
applied uniformly everywhere [ to $(\lambda ,\theta $ ) to components of a
vector.]. The nontrivial relation between the renormalization constants of all
sources for fields and composite operators are shown to arize from the fact that 
they belonged to supermultiplets. A ``nonrenormalization theorem" was proved
for the symmetry breaking piece of the action  $S_1$ (which generates
sources for fields \underline{and} composite operators, gauge-fixing term
and the ghost term ) that $S_1$ is left form invariant under
renormalization .

The results mentioned above from Ref. \cite{D49}, were not ``derived"
directly but shown to hold using the correspondence between the 
superspace formulation and the usual Yang-Mills theory as given in
Eq. \ref{29} and \ref{23}. One of the purposes of the present work is to
derive them  and show how simply they arize once the WT identity for 
generating functional $\Gamma $ for proper vertices is formulated. The task
of obtaining this WT identity for $\Gamma $ taken up in the next section.

\section{DERIVATION OF WT IDENTITIES FOR PROPER VERTICES}

The purpose of this section is to derive the WT identity for the generating
functional for proper vertices. The procedure is generally straight
forward.
We derive the WT identity for $\bar{W} $ ( or  rather a related quantity
$ \bar{W} ^\prime $ of Eq. \ref{31} below); and use Legendre transform of $
\bar{Z} $ to obtain the desired result. Only a few subtleties not
encountered in the treatment in the ordinary space have to be dealt with
carefully.

We shall find it convenient to deal not with $\bar{W} $ of Eq. \ref{29}
but rather with
\begin{equation}
\bar{W}^\prime \equiv \int [dK^4][dK^4_{,\theta }] \bar{W} [\bar{K} ,\bar{L} ,t]
\label{31}
\end{equation}
as this is the quantity related to generating functional of ordinary Yang-Mills theory via an equation like
\ref{29}.

Finally, we shall derive the WT identity satisfied by $\bar{W} ^\prime [\bar{K},
\bar{L} ,t]$. The derivation here follows the method of Ref. 9 as it is briefer.
We shall find it convenient to use the form of $ \bar{W} ^\prime [\bar{K}, \bar{L} ,t]$
that is obtained by integrating out the auxiliary fields. We obtain $\bar{W} ^\prime $,
 after these integrations as,
 \begin{eqnarray}
 \bar{W} ^\prime =\int \left[ dA dc_5 d \zeta\right] \exp\Big[ i\int d^4x
 \left\{ -\frac{1}{4}F_{\mu\nu}F^{\mu\nu} +(K^\mu - \partial ^\mu
 \zeta )Dc_5 -\frac{1}{2} K^5 gfc_5c_5 \nonumber\right. && \\
 \left. -\frac{ \eta_0}{2}( \partial\cdot A +t)^2 + K^\mu_{, \theta }
 A_\mu + K^5_{ ,\theta}c_5 +t_{, \theta } \zeta-K^\mu
 L^\mu +\frac{1}{2}
 K^5L^4 -\frac{1}{2}(L^5)^2 \right\}\Big] &&\label{32}
 \end{eqnarray}
We then proceed in the following steps as in Ref. 9.
(i) Evaluate $\frac{ \partial \bar{W} ^\prime}{\partial \theta }$
explicitly using Eq.~ \ref{32} as \\

\begin{eqnarray}
\frac{\partial W^\prime}{\partial\theta} = K^\mu_{,\theta}< Dc_5> -\frac{1}{2}
K^5_{,\theta}< gfc_5c_5> -<\eta_0(\partial\cdot A +t)>t_{,\theta} \nonumber \\ +i\frac{\partial}{\partial\theta}
\left [ -K^\mu L^\mu +\frac{1}{2} K^5L^4 -\frac{1}{2}(L^5)^2\right]\bar{W}^\prime
\label{3200}
\end{eqnarray}

(ii) Obtain a WT identity using BRS invariance of the action appearing on
the right hand side of Eq. \ref{32}.

\begin{equation}
K^\mu_{,\theta}<Dc_5> -\frac{1}{2}K^5_{,\theta}<gfc_5c_5> -<\eta_0(\partial\cdot A +t)> t_{,\theta} =0
\label{3300}
\end{equation}

(iii) We next simplify the result for $ \frac{ \partial \bar{W} ^\prime }{ \partial \theta }
 $ in Eq. \ref{3200} using Eq. \ref{3300} to obtain the WT identity.
\begin{equation}
\frac{ \partial \bar{W} ^\prime }{ \partial \theta } = -i\frac{
\partial }{ \partial \theta } \left[ K^\mu L^\mu
-\frac{1}{2}K^5 L^4 + \frac{1}{2}(L^5)^2\right]\bar{W}^\prime
\label{33}
\end{equation}
Note that Eq. \ref{33} is consistent with the simple result $\frac{ \partial 
\bar{W} }{\partial \theta } =0 $ of Ref. \cite{osp} at $ L=0$.

The generating functional of constructed Green's functions $\bar{Z} [\bar{K}, \bar{L},t]
$ is given by 
\begin{equation}
\bar{Z} [ \bar{K} , \bar{L} ,t] = -i\mbox{ ln }\bar{W} ^\prime [ \bar{K} .\bar{L}
,t] \label{34}
\end{equation}
[Here we note that $ \bar{W} ^\prime $ is a function of $\lambda $
and $\theta $ (through sources). The logarithm of a function of a
Grassmannian defined as  $ \log(a + b \theta ) = \log [ a(1+ \frac{ b}{a} \theta )]
= \log (ae^{ \frac{ b}{a} \theta}) = \log a + \frac{ b}{a}\theta $. Similarly, $\log(a+b \theta 
+c \lambda )] = \log a + \frac{ b}{a} \theta + \frac{ c}{a} \lambda + \frac{ bc}{a} \lambda \theta $ 
etc ]

Next we define the generating functional for proper vertices
\begin{equation}
\Gamma = \bar{Z} - \sum_{S_i}\int S_i(x) \frac{ \delta \bar{Z} }{\delta
S_i(x)} d^4x \label{35}
\end{equation}
Where $ S_i(x)$ generally denotes all sources ( as function of $x$ only)
that appear for elementary fields viz. $ S_i(x)\equiv \left\{ K^\mu ,K^\mu_{, \theta }
,K^5,K^5_{,\theta }, L^\mu , L^5, L^4, t, t_{,\theta}\right\}$. From now on
we shall find it convenient to drop explicit $\lambda $ occuring in $\bar{Z} $
and $\Gamma $ though it is not necessary. Then Eq. \ref{35} can
alternatively be written as
\begin{equation}
\Gamma = \bar{Z} - \sum_{S_i(\bar{x} )}\int S_i(\bar{x}) \frac{ \delta^\prime  
\bar{Z} }{\delta
S_i(\bar{x})} d^4x \label{36}
\end{equation}
Here, now, $S_i(\bar{x} )$ are function of $x$ and $\theta $ and $\frac{ \delta 
^\prime }{\delta S_i( \bar{x} )}$ denotes differentiating $S_i( \bar{x} ) $
taking into account only the \underline{explicit} $ S_i(\bar{x} )$
dependence. [ e.g. while differentiating with respect to $K^\mu_{, \theta
}$, we don't consider the $K^\mu_{, \theta } $ dependence in $K^\mu( \bar{x} )
$]. This is easily verified using the identities such as
\begin{eqnarray}
\frac{ \delta \bar{Z} }{\delta K^\mu(x)}& =& \frac{ \delta ^\prime \bar{Z}
}{\delta K^\mu( \bar{x} )}\nonumber\\
\frac{ \delta \bar{Z} }{\delta K^\mu_{, \theta }(x)}& =& \frac{ \delta ^\prime \bar{Z}
}{\delta K^\mu_{, \theta }( \bar{x} )} - \theta \frac{ \delta ^\prime
Z}{\delta K^\mu( \bar{x} )}
\label{37}
 \end{eqnarray} 
$\Gamma $ defined in Eq. \ref{35} is a functional of expectation values
$\frac{ \delta \bar{Z} }{\delta S_i(x)}$ as usual. By rearrangement of
variables, it can also re-expressed as a functional of variables $\frac{ \delta 
^\prime \bar{Z} }{\delta S_i( \bar{x} )}$. Thus $\Gamma $ can be looked
upon as function of
\begin{equation}
\Gamma = \Gamma \left [ <A_\mu (\bar{x}) >,<A_{\mu , \theta }>,<c_5
(\bar{x}) >, <c_{5,\theta }>,< A_{\mu,\lambda }^\prime >, <c_{5,\lambda }>,
<c_{4, \lambda }>,<\zeta>,<\zeta_{, \theta }>\right ]
\label{38}
\end{equation}

Now, just as $\bar{W} ^\prime $ had no \underline{explicit} dependence on
$\theta $, when looked up on as a functional of $\bar{K} (\bar{x}), K_{, \theta} 
(\bar{x}), \bar{L} (\bar{x}) , t (\bar{x}) , t_{, \theta } (\bar{x}) $
[i.e. all its $ \theta $-dependence arose from its dependence on sources
$ \bar{K} (\bar{x}) , t (\bar{x}) $], it can be shown [ See Appendix]
that $\Gamma $ seen as functional of variables listed in Eq. \ref{38}
has no \underline{explicit} dependence on $\theta $ either.

Now from Eq. \ref{33}, $\bar{Z} $ can be shown to satisfy
\begin{equation}
\frac{ \partial \bar{Z} }{\partial \theta } = -\frac{ \partial }{\partial
\theta } \left [ K^\mu L_\mu -\frac{ 1}{2} K^5 L^4 +\frac{ 1}{2}(L^5)^2\right ]
\label{39}
\end{equation}

This then leads, straightforwardly, using Eq. \ref{36} to the WT identity
for
$\Gamma $.
\begin{equation}
\frac{ \partial \Gamma }{ \partial \theta } = L^\mu K_{\mu, \theta }
-\frac{ 1}{2} L^4 K^5_{, \theta }
\label{310}
\end{equation}

Here, we understand that $ L^\mu, L^4, K^\mu_{, \theta }$ and $K^5_{, \theta
}$ are to be expressed back in terms of expectation values. Here, we note
that
the arguments of $\Gamma $ are, as mentioned earlier,
\begin{eqnarray}
\left < A_\mu (\bar{x}) \right > = \frac{ \delta ^\prime \bar{Z} }{\delta
K_{\mu,\theta} (x)} &;\;\;& \left < A_{\mu, \theta} (x)\right > = \frac{ \delta
^\prime \bar{Z} }{\delta K_\mu (\bar{x}) }\nonumber \\
\left < A^\prime _{\mu,\lambda }\right > = \left < A_{\mu, \lambda } +
\partial _\mu \zeta \right > &\equiv & -\frac{ \delta ^\prime \bar{Z} }{
\delta L^\mu}
\label{311}
\end{eqnarray}
\hspace{ 4.5in} etc.

Now, we note that (we drop brackets around expectation values)
\begin{eqnarray}
L^\mu =- \frac{ \delta \Gamma }{\delta A_{\mu,\lambda} }; &\;\;\; & K^\mu_{, \theta }
= - \frac{ \delta \Gamma }{ \delta A_\mu}\nonumber
\\
L^4 = -\frac{ \delta \Gamma }{\delta c_{4,\lambda} }; &\;\;\; & K^5_{, \theta } =
\frac{ \delta \Gamma }{ \delta c_5}
\label{312}
\end{eqnarray}
We thus have the WT identity 
\begin{equation} 
\frac{ \partial \Gamma }{ \partial \theta } = \frac{ \delta \Gamma }{\delta
A_{\mu,\lambda }}\frac{ \delta \Gamma }{ \delta A_\mu} + \frac{ 1}{2} \frac{
\delta \Gamma }{\delta c_{4, \lambda }} \frac{ \delta \Gamma }{ \delta c_5}
\label{313}
\end{equation}

We can deal with the renormalization of gauge theories from this equation
itself. However we prefer to use a somewhat alternate approach using a
somewhat unusual change of variables.

Firstly, we denote collectively
\begin{eqnarray}
\Phi_i (\bar{x}) \equiv && \left \{ <A_\mu (\bar{x}) >, <c_5 (\bar{x})
>\right\}\label{314}\\
\left < A_{i, \lambda }^\prime \right > \equiv && \left \{ < A^\prime
_{\mu,\lambda }>, <c_{4,\lambda }> \right \} \label{315}
\end{eqnarray}
We now want to calculate $\frac{ \partial }{ \partial \theta } \Phi _i (\bar{x})
$. We note
\begin{eqnarray}
\frac{ \partial }{ \partial \theta }\Phi_\mu (\bar{x})  &=&
\frac{ \partial }{ \partial \theta }\left[\frac{ \delta ^\prime
\bar{Z}  }{ \delta K^\mu_{,\theta} }\right] \nonumber \\
&=& \frac{ \partial}{ \partial \theta }\left[ \frac{ \delta 
\bar{Z}  }{ \delta K^\mu_{,\theta}  }+ \theta \frac{ \delta 
\bar{Z}  }{ \delta K^\mu}\right]\nonumber \\
&=& \frac{ \delta }{ \delta K^\mu_{,\theta} }\left[\frac{ \partial
\bar{Z} }{ \partial \theta }\right] +\theta \frac{ \delta
}{ \delta K^\mu }\left[\frac{ \partial \bar{Z} }{ \delta \theta } \right]+
\frac{ \delta \bar{Z} }{ \delta K^\mu}\nonumber \\
&=& -L^\mu + <A_{\mu , \theta }> \label{316}\\
\mbox{Similarly      }\nonumber \\
 \frac{ \partial }{ \partial \theta } \Phi_5 (\bar{x}) &=&
-\frac{1}{2}L^4 + <c_{5, \theta }>\label{3165} \\
\frac{ \partial }{\partial\theta}<\zeta> &=& <\zeta_{,\theta}>
\label{317}
\end{eqnarray}
Now if we define 
\begin{eqnarray}
\Phi_\mu^\prime (\bar{x})  &=&\Phi_\mu (\bar{x}) + \theta L^\mu\nonumber\\
\Phi_5^\prime (\bar{x})  &=&\Phi_ 5(\bar{x}) + \theta \frac{1}{2}L^4
\label{318}
\end{eqnarray}
Then, Eq.~\ref{316} and \ref{3165} can be written in nice form
\begin{eqnarray}
\frac{ \partial }{ \partial \theta }\Phi^\prime _\mu (\bar{x}) &=&<
A_{\mu, \theta }> \nonumber \\
\frac{ \partial }{ \partial \theta }\Phi^\prime _5 (\bar{x}) &=&<
c_{5, \theta }> \label{319}
\end{eqnarray}

In terms of these redefined variables the WT identity for $\Gamma $ reads
\begin{eqnarray}
\frac{ \partial }{ \partial \theta }\Gamma\left[\Phi_\mu^\prime
(\bar{x})- \theta L^\mu ,\Phi_5^\prime (\bar{x}) -\frac{ \theta
}{2}L^4 ,<A_{i, \theta }(\bar{x})>, <A^\prime _{i,
\lambda}(\bar{x})>, <\zeta (\bar{x})>,<\zeta_{, \theta }> \right]\nonumber && \\
=L^\mu K^\mu_{, \theta
}-\frac{1}{2}L^4K^5_{, \theta }&&\label{320}
\end{eqnarray}
i.e. 
\begin{eqnarray} 
\frac{\partial}{\partial\theta}\left\{
\Gamma\left[\Phi_\mu^\prime,\Phi^\prime_5,<A_{i,\theta}>,<A_{i,\lambda}^\prime>,<\zeta (x)>,<\zeta_{,\theta}>\right ] 
-\theta L^\mu\frac{\delta\Gamma}{\delta\Phi_\mu} -\frac{\theta}{2}L^4\frac{\delta\Gamma}{\delta
c_5}\right\}\nonumber \\
= L^\mu K^\mu_{,\theta}-\frac{1}{2}L^4K^5_{,\theta}
\label{321}
\end{eqnarray}
Now using the expression \ref{312} for $L^\mu,K^\mu_{,\theta} $ etc, we
obtain
\begin{equation}
\frac{ \partial }{ \partial \theta }\Gamma\left[\Phi_\mu^\prime
(\bar{x}),\Phi_5^\prime (\bar{x}) 
 , <A_{i, \theta } (\bar{x})> , <A^\prime _{i,
\lambda }(\bar{x})>, <\zeta (\bar{x})>,<\zeta_{, \theta }> \right]=0
\label{322}
\end{equation}
Noting the relation \ref{319} for $<A_{i,\theta }>$, we need not write
these as separate variables. Then the WT identity for $\Gamma $ can be
compactly written as,
\begin{equation}
\frac{ \partial }{ \partial \theta }\Gamma\left[\Phi_i^\prime 
(\bar{x}), 
<A^\prime_{i,\lambda }(\bar{x})>, <\zeta (\bar{x})> \right]=0
\label{323}
\end{equation}
[ We should clarify that while our attempt to put the WT identity in the
neat form of Eq.~\ref{323} by change of variables of \ref{318} may seem a
bit artificial, it is not quite so. As we shall see in the next section,
under renormalization, not only the old variables $\Phi_i$ but {\it also }
the new variables $\Phi_i^\prime$ of \ref{318} become multiplicatively
renormalized. More on this later. Also the new variables are more natural
since it is the new variables which directly contain $<A_\mu(x)>$ and $<A_{\mu,\theta}>$
as components: $\Phi^\prime(\bar{x})= <A_\mu(x)> +\theta<A_{\mu,\theta}>+\cdots $]

In addition to Eq. \ref{323}, there is further restriction on the expectation value
$< A^\prime _{\mu,\lambda }>$. This is seen, for example, as follows:
\begin{equation}
\frac{ \partial }{ \partial \theta } < A^\prime _{\mu, \lambda }> = \frac{ \partial }
{\partial \theta }\frac{ \delta Z}{\delta L^\mu} = \frac{ \delta }{\delta L^\mu}
\frac{ \partial Z}{\partial \theta } =- K^\mu_{, \theta} (x) = \frac{ \delta \Gamma 
}{ \delta A_\mu} \label{324}
\end{equation}
In general it is easy to see that
\begin{equation}
\frac{ \partial }{\partial \theta }< A^\prime _{i,\lambda }> = \eta_i \frac{ 
\delta \Gamma }{\delta A_i} \label{325}
\end{equation}
with $\eta_\mu =1, \eta_4 = \frac{ 1}{2}, \eta_5 = 0$. [ The index `$i$' on the
right hand side of the Eq. \ref{325} in not summed over]. Equations \ref{323} and \ref{325}
constitute the basis for discussion of renormalization.

\section{ SOLUTION TO WT IDENTITIES AND RENORMALIZATION}
In previous section, we had arrived at the WT identity for proper vertices 
\begin{eqnarray}
\frac{ \partial }{ \partial \theta } \Gamma \left[ \Phi_i^\prime ,
<A^\prime_{i , \lambda }>,<\zeta>\right] &=& 0 \label{41}\\
\mbox{with the subsidiary constraint\hspace{3in}} &&\nonumber \\
\frac{ \partial }{ \partial \theta }<A^\prime_{i, \lambda }> &=&\eta_i \frac{ \delta 
\Gamma}{ \delta A^{i}}\label{42}
\end{eqnarray}
Before we proceed with the discussion of renormalization of gauge theories,
we shall draw attention to the advantages of our formulation.

(1) Our formulation does not contain any composite operators explicitly, it
contains only the `` elementary fields" and as such our discussion of
renormalization centers on only the renormalization of these elementary
fields .\\
(2) The form of the WT identity in \ref{41} is extremely simple, compared to
the earlier formulations \cite{lee,pr1}.
In particular, the nilpotent operator ${\cal G}$ of Ref. 3 or its analogies are 
related by a very simple operator  $ \frac{ \partial }{ \partial \theta }
$.
More importantly this latter operator is independent of $g$ and fields
unlike the nilpotent operator ${\cal G} $ \cite{lee}.
\begin{equation}
{\cal G} = \int \left [ {\cal D} _\mu c \frac{ \delta }{\delta A_\mu}
-\frac{ 1}{2} g_0 fcc \frac{ \delta }{\delta c} + \frac{ \delta
\tilde{S}}{\delta A^\alpha _\mu } \frac{ \delta }{ \delta \kappa^\alpha_\mu} +
\frac{ \delta \tilde{S}}{\delta c^\alpha } \frac{ \delta }{ \delta l^\alpha
}\right ]\label{43}
\end{equation}
(used in the original discussion of renormalizability) which depends in a
complicated fashion on the coupling constant $g$ and the fields and
sources.
This simplifies greatly and makes it very transparent the kind of
renormalization allowed by \ref{41} and \ref{42} as seen below in (3).

(3)In the original formulation of renormalization using BRS \cite{lee}
it does require some work to prove that rescaling of $g$ is an allowed renormalization.
This requires showing that the operator $ O = g \frac{ \partial \tilde{S}}{\partial
g} $ is a solution of 
\begin{equation}
{\cal G } O = 0 \label{44}
\end{equation}
To deduce this form 
\begin{equation}
{\cal G}\tilde{S} =0 \label{45}
\end{equation}
requires some work because ${\cal G}$ itself depends on $g$ and ${\cal G }( g 
\frac{ \partial \tilde{S}}{\partial g}) =0$ can't be deduced by
differentiating Eq. \ref{45} with respect to $g$ . A similar statement holds
about alternate formulations \cite{pr1}. In the present formulation,
however,
the operator $g \frac{ \partial }{\partial g}$ commutes with $\frac{ \partial }{\partial \theta }$.
Hence if $\Gamma $ satisfies \ref{41}[ and \ref{42} ], then $\Gamma + \epsilon g 
\frac{ \partial }{\partial g} \Gamma $ also satisfies the same equations is
trivially established. In fact, it is easy to see that as the operator $\frac{ 
\partial }{\partial \theta }$ is independent of fields also, any
multiplicative renormalization of $A_\mu , c_5 ,A_{\mu,\theta },c_{5,\theta },
c_{4,\theta }$ etc are also compatible with Eq. \ref{41}, the form of Eq.~
\ref{42} determine the restrictions to be placed on these rescalings.

Having brought out the advantages of our formulations, we now proceed to
determine the renormalization transformations and show that they agree with
those of Ref. 14

At first, we shall assume a multiplicative renormalization  on superfields,
$\lambda ,\theta $ and coupling constants and seek a solution that is
consistent with it. The justification for this procedure will be given at
the end of this section.

We imagine rescalings
\begin{eqnarray}
\lambda = Z_ \lambda \lambda ^R;& \; &\,\, \theta = Z_ \theta \theta ^R
\nonumber \\
A_\mu =Z^{\frac{1}{2}}A_\mu^R; & \; & c_4= Z^{\frac{1}{2}}Z_{(4)}c_4^R;\,\,
c_5 = Z^{\frac{1}{2}}Z_{(5)}c_5^R \label{46}
\end{eqnarray}
All such rescaling are trivially compatible with Eq.~ \ref{41}. Thus, if
they are also to satisfy Eq.~ \ref{42}, all that is needed is
\begin{equation}
Z_ \lambda Z_ \theta = Z = Z_{(4)}Z_{(5)}Z
\label{47}
\end{equation}
Now define
\begin{equation}
\frac{Z_\theta}{Z_{(4)}} = \tilde{Z}\label{48}
\end{equation}
Then equation \ref{47} yields 
\begin{equation}
Z_ \lambda = \frac{ Z}{ \tilde{Z} Z_{(4)}} \label{49}
\end{equation}
We relabel 
\begin{equation}
Z_{(4)} = Z^{(5)}\;\;\;\mbox{and  }\;\;\; Z_{(5)}= Z^{(4)} \label{410}
\end{equation}
and parameterize
\begin{equation}
Z^{(5)} = Z_1 \frac{ Z^{\frac{ 1}{2}}}{\tilde{Z}} \label{411}
\end{equation}
Then we have the following set of transformations
\begin{eqnarray}
Z_ \lambda &=& \frac {Z}{\tilde{Z} Z^{(5)}} = \frac{ Z}{\tilde{Z} }Z^{(4)} =
Z^{\frac{ 1}{2}}Z_1^{-1}\label{412a}\\
Z_ \theta &=& \tilde{Z} Z^{(5)}\label{412b} \\
A_\mu (\bar{x}) &=& Z^{\frac{ 1}{2}} A_\mu ^R (\bar{x}^R) ;\;\;\;\; c_4 (\bar{x}) =
Z^{\frac{ 1}{2}} Z_{(4)} c_4^R (\bar{x}^R); \;\;\;\;c_5 (\bar{x}) = Z^{\frac{ 1}{2}}
Z_{(5)}c_5^R (\bar{x}^R)\label{412c}
\end{eqnarray}
In view of Eq. \ref{42}, it follows that as $A_\mu $, and $\theta $ multiplicatively
renormalized , so is $A^\prime _{\mu, \lambda } = A_{\mu,\lambda } -\partial _\mu 
\zeta $. Hence the renormalization of $\zeta$ is the same as that of $A_{\mu,\lambda }
$ viz.
\begin{equation}
A_{\mu, \lambda } = Z^{\frac{ 1}{2}} Z_{\lambda }^{-1} A_{\mu,\lambda }^R =
Z_1 A^R_{\mu, \lambda }\label{412d}
\end{equation}
Hence
\begin{equation}
\zeta (\bar{x}) = Z_1 \zeta^R (\bar{x}^R) \label{412e}
\end{equation}
In view of the fact that $ K^\mu_{, \theta }, t_{, \theta }$ are sources for $A_\mu$
and $\zeta $ and must transform contragradriently, we further have
\begin{equation}
K^i (\bar{x}) = Z_1 K^{iR} (\bar{x}^R);\;\;\; t (\bar{x}) = Z^{\frac{ 1}{2}}
t^R (\bar{x}^R) \label{412f}
\end{equation}
The equations \ref{412a}-\ref{412f} in fact represents the renormalization 
transformations \footnote{ The renormalization transformation of
Eq. \ref{412a}- \ref{412f} differ slightly from those written down in Ref. 14 in that the
renormalization of $\lambda $ is different here. In Ref. 14, no sources were introduced for $
 A_{i, \lambda }$; and hence renormalization of $ A_{i, \lambda }$ ( and therefore
 of $\lambda $) were arbitrary there. $Z_ \lambda $ was fixed by convention
in
Ref. 14 so that $\lambda $ and $\theta $ renormalize symmetrically. But
this assumption is not necessary there. It is only in this work that $Z_ \lambda
$ is fixed uniquely and is given by \ref{412a}.}
written down in Ref. 14. And that they represent solution to \ref{41} and
\ref{42} has been seen almost trivially.

This then would complete the proof of the superfield renormalization
transformation except that we have assumption of multiplicative
renormalization of superfields as a whole. We shall now tern to justifying
this.

In \ref{41} we expect $\Gamma = \Gamma ^R$ . Hence the only freedom
we have is for transforming $\theta $ such that
\begin{equation}
\frac{ \partial }{ \partial \theta } = Z_ \theta ^{-1} \frac{\partial
}{\partial \theta ^R} \label{413}
\end{equation}
for then Eq. \ref{41} will read $ \frac{ \partial}{\partial \theta ^R} \Gamma 
^R =0 $ i.e. it will remain form invariant.
Now $ A_\mu(x), c_5(x)$ are all dimension 1 operators and hence cannot mix 
with any other field on account of dimension, Lorentz transformation and 
 ghost number. Hence they must be multiplicatively renormalizable.
[ Hence we are talking only of $\theta , \lambda $ independent piece of
$A_\mu (\bar{x}) $ and $c_5 (\bar{x})  $].
Taking into account Eq. \ref{413}, Eq. \ref{42} implies that $A^\prime 
_{i, \lambda }$ should be multiplicatively renormalizable. Writing
\begin{equation} 
A_{i, \lambda }^\prime (x) = A^{\prime R}_{i, \lambda }Z^{-1}_i Z_ \theta 
\label{414}
\end{equation}
we note
\begin{equation}
\frac{ \delta }{ \delta A^\prime_{i, \lambda }} = Z_i Z_ \theta ^{-1} \frac{ \delta
}{\delta A^{\prime R}_{i, \lambda }} \label{415}
\end{equation}

 Now note that using \ref{42} ,
\begin{equation}
\frac{ \partial \Gamma }{\partial \theta } = \frac{ \delta \Gamma }{\delta
\Phi ^\prime _i} \Phi ^\prime _{i, \theta } + \frac{ \delta \Gamma }{\delta
<\zeta> }<\zeta_{,\theta }>+ \frac{ \delta \Gamma }{ \delta A^\prime_{i, \lambda }}
\frac{ \delta \Gamma }{\delta A_i} \eta_i
\label{4155}
\end{equation}
Now $\frac{ \partial \Gamma }{ \partial \theta }$ gets renormalized by $Z^{-1}_
\theta $; and so does
\begin{equation}
\frac{ \delta \Gamma }{\delta A^\prime_{i, \lambda }} \frac{ \delta \Gamma }{\delta
A_i}\eta _i = Z_i Z_ \theta ^{-1} \frac{ \delta \Gamma ^R}{\delta A^{\prime R}_{i,
\lambda }} Z_i^{-1}\frac{ \delta \Gamma }{ \delta A^R_{i}}\eta_i =
Z_ \theta ^{-1} \frac{ \delta \Gamma ^R}{ \delta A^{\prime R}_{i, \lambda }} \frac{ \delta 
\Gamma }{\delta A_i^R} \eta_i
\label{416}
\end{equation}
Hence for  $ \frac{ \delta \Gamma }{\delta \Phi _i} \Phi _{i,\theta }
$ also we must have 
\begin{equation}
\frac{ \delta \Gamma }{\delta <\zeta>} <\zeta_{, \theta }> +\frac{ \delta 
\Gamma }{\delta \Phi _i} \Phi _{i, \theta } = Z_ \theta ^{-1} \left [
\frac{ \delta \Gamma ^R}{\delta \Phi ^R_i}( \Phi _{i, \theta })^R + \frac{ 
\delta \Gamma ^R}{\delta <\zeta^R> }< \zeta_{, \theta }>^R \right ]
\label{417}
\end{equation}
This implies that $ \Phi _i(x)$ and $ \Phi _{i, \theta } (x)$ get renormalized
in such a fashion that
\begin{eqnarray}
\Phi _i (x, \theta ) = \Phi _i(x) + \theta \Phi _{i, \theta }(x) &=&
Z_{\Phi }\left ( \Phi ^R_i(x) + \theta ^R (\Phi _{i, \theta })^R(x) \right
)\nonumber \\
&\equiv & Z_{\Phi }\left [ \Phi ^R_i(x) + \theta ^R \Phi ^R_{i, \theta
^R}(x) \right ] \nonumber \\
&=& Z_{\Phi} \Phi ^R_i(x, \theta ^R) \label{418}
\end{eqnarray}
This justifies the assumption that $\Phi _i(x, \theta )$ is multiplicatively
renormalized as a whole. Now note that the $\Gamma $ in Eqs \ref{41} and 
\ref{42} is $\lambda $ independent as $\lambda$ has been set to zero. The
only reference to $\lambda $ comes from $A^\prime _{i, \lambda } $. Hence
we can always \underline{choose} $Z_ \lambda $ such that $A_{i, \lambda }$
is renormalized as
\begin{equation}
A_{i, \lambda } (x) = Z_i Z_ \lambda ^{-1} A_{i, \lambda }^R (x)
\label{419}
\end{equation}
This justifies the assumptions made earlier.

\section{ CONCLUSIONS}

    In conclusion, we have, in this work, generalized and carried forward our
 earlier work on superspace formulation of gauge theories by introducing the
 generating functional for proper vertices in this formulation and have shown
 that the WT identities can be cast in the same simple form [as for $\bar{ W }$]
 viz. $\frac{d\Gamma}{d\theta}=0$ . We have further shown how this simplifies the discussion
 of superspace renormalization of gauge theories and how it leads to the
 supermultiplet renormalization [ and hence other elegant results ] that were
 formulated in Ref.14. The simplification has been due to the fact that the
 nilpotent operator $\frac{d}{d\theta} $ occuring in the WT identities for $ \Gamma $ is independent
 of the variables such as fields ans coupling constants.
\newpage
\appendix
\section{}
 
Let $S_i (\bar{x}) $ denotes collectively all ``whole" sources [ such as
$\bar{K} (\bar{x}) , \bar{K}_{, \theta } (\bar{x}) \cdots$ etc]. Then,
\begin{equation}
\bar{Z} = \bar{Z} [ S_i (\bar{x}) ] \label{a1}
\end{equation}
and has no explicit dependence on $\theta $. We define the modified
expectation values denoted collectively by $\Psi_i$
\begin{equation}
\Psi_j (\bar{x}) \equiv \frac{ \delta ^\prime \bar{Z} [S_i]}{\delta S_j
(\bar{x}) } \label{a2}
\end{equation}
Thus $\Psi_j (\bar{x}) $ when expressed as a functional of ``whole" sources
$ S_i (\bar{x}) $ has no explicit $\theta $ dependence.
Now
\begin{equation}
\frac{ \partial }{\partial \theta }\Psi_j (\bar{x})\left |_{\Psi_j (\bar{x}) }
\equiv 0 = \int d^4x \frac{ \partial }{ \partial \theta
}S_k(\bar{y})|_{\Psi_j} \frac{ {\delta ^\prime }^2 \bar{Z} [S]}{\delta S_k(
\bar{y})\delta S_j (\bar{x}) }\right . \label{a3}
\end{equation}
Assuming $ \frac{ { \delta ^\prime }^2 Z}{ \delta S_k(\bar{y}) \delta S_j (\bar{x})}
$ to be an invertible matrix as is needed for the invertibility \& to
obtain $S_i (\bar{x}) $ in terms of $\Psi_j$, we hence obtain
\begin{equation}
\frac{ \partial }{\partial \theta } S_k(\bar{y})\left |_{\Psi_j =0}\right .
\label{a4}
\end{equation}
hence when $S_k$ are expressed as functionals of $\Psi_j$, they
contain no explicit $\theta $ dependence.
\begin{equation}
S_k = S_k [\Psi_j] \label{a5}
\end{equation}
We substitute this expression for $S_k$ in
\begin{equation}
\Gamma = \bar{Z} - \sum_i\int S_i (\bar{x}) \frac{ \delta ^\prime \bar{Z}
}{ \delta S_i (\bar{x}) } d^4x
\label{a6}
\end{equation}
to obtain
\begin{equation}
\Gamma = \Gamma [ \Psi_j]
\label{a7}
\end{equation}
i.e. $\Gamma $ has no explicit $\theta $ dependence when expressed as
functional of $\Psi_j$.

\end{document}